\documentclass{iaus}
\usepackage{graphicx}

\title[S272.~~Active OB stars] 
{Infrared properties of Active OB stars \\ in the Magellanic Clouds from
  the\\ Spitzer SAGE Survey}

\author[Bonanos et al.]   
{A.Z. Bonanos$^1$,
  D.J. Lennon, D.L. Massa, M. Sewilo, F. Koehlinger, N. Panagia,
  J.Th. van Loon, C.J. Evans, L.J. Smith, M. Meixner, \\ K. Gordon \and
  the SAGE teams}

\affiliation{$^1$National Observatory of Athens, IAA, \\ I. Metaxa \&
  Vas. Pavlou Street, Palaia Penteli GR-15236, Greece\\ email: {\tt
    bonanos@astro.noa.gr}}

\pubyear{2010}
\volume{xxx}  
\pagerange{119--126}
\jname{Active OB stars: structure, evolution, mass loss and critical limits}
\editors{C. Neiner, G. Wade, G. Meynet \& G. Peters, eds.}
\begin{document}

\maketitle

\begin{abstract}

  We present a study of the infrared properties of 4922
  spectroscopically confirmed massive stars in the Large and Small
  Magellanic Clouds, focusing on the active OB star
  population. Besides OB stars, our sample includes yellow and red
  supergiants, Wolf-Rayet stars, Luminous Blue Variables (LBVs) and
  supergiant B[e] stars. We detect a distinct Be star sequence,
  displaced to the red, and find a higher fraction of Oe and Be stars
  among O and early-B stars in the SMC, respectively, when compared to
  the LMC, and that the SMC Be stars occur at higher luminosities. We
  also find photometric variability among the active OB population and
  evidence for transitions of Be stars to B stars and vice versa. We
  furthermore confirm the presence of dust around all the supergiant
  B[e] stars in our sample, finding the shape of their spectral energy
  distributions (SEDs) to be very similar, in contrast to the variety
  of SED shapes among the spectrally variable LBVs.

\end{abstract}

\firstsection 
\section{Introduction}

The {\em Spitzer Space Telescope} Legacy Surveys SAGE (``Surveying the
Agents of a Galaxy's Evolution'', \cite[Meixner et
al. 2006)]{Meixner06} and SAGE-SMC \cite[(Gordon et
al. 2010)]{Gordon10} have for the first time made possible a
comparative study of the infrared properties of massive stars at a
range of metallicities, by imaging both the Large and Small Magellanic
Clouds (LMC and SMC). In \cite[Bonanos et al. (2009]{Bonanos09}, Paper
I) and \cite[Bonanos et al. (2010]{Bonanos10}, Paper II), we presented
infrared properties of massive stars in the LMC and SMC, which we
summarize below. The motivation was threefold: (a) to use the infrared
excesses of massive stars to probe their winds, circumstellar gas and
dust, (b) to provide a template for studies of other, more distant,
galaxies, and (c) to investigate the dependence of the infrared
properties on metallicity. Papers I and II were the first major
compilations of accurate spectral types and multi-band photometry from
0.3$-$24 $\mu$m for massive stars in any galaxy, increasing by an
order of magnitude the number of massive stars for which mid-infrared
photometry was available.

Infrared excess in hot massive stars is primarily due to free-free
emission from their ionized, line driven, stellar winds. \cite[Panagia
\& Felli (1975)]{Panagia75} and \cite[Wright \& Barlow
(1975)]{Wright75} first computed the free-free emission from ionized
envelopes of hot massive stars, as a function of the mass-loss rate
($\dot{M}$) and the terminal velocity of the wind ($v_{\infty}$). The
properties of massive stars, and in particular their stellar winds
(which affect their evolution) are expected to depend on metallicity
($Z$). For example, \cite[Mokiem et al. (2007)]{Mokiem07a} found
empirically that mass-loss rates scale as $\dot{M} \sim Z^{0.83\pm
  0.16}$, in good agreement with theoretical predictions (\cite[Vink
et al. 2001]{Vink01}). The expectation, therefore, is that the
infrared excesses of OB stars in the SMC should be lower than in the
LMC, given that $\dot{M}$ is lower in the SMC. Furthermore, there is
strong evidence that the fraction of classical Be stars among B-type
stars is higher at lower metallicity \cite[(Martayan et
al. 2007)]{Martayan07b}.  \cite[Grebel et al. (1992)]{Grebel92} were
the first to find evidence for this, by showing that the cluster
NGC\,330 in the SMC has the largest fraction of Be stars of any known
cluster in the Galaxy, LMC or SMC. More recent spectroscopic surveys
\cite[(Martayan et al. 2010)]{Martayan10} have reinforced this
result. We are also interested in quantifying the global dependence of
the Be star fraction on metallicity.  The incidence of Be/X-ray
binaries is also much higher in the SMC than in the LMC \cite[(Liu et
al. 2005)]{Liu05}, while the incidence of Wolf-Rayet (WR) stars is
much lower; therefore, a comparison of infrared excesses for these
objects is also of interest.

\firstsection
\section{Spectral type and Photometric Catalogs}

We have compiled catalogs of massive stars with known spectral types
in both the LMC and SMC from the literature. We then cross-matched the
stars in the SAGE and SAGE-SMC databases, after incorporating optical
and near-infrared photometry from recent surveys of the Magellanic
Clouds. The resulting photometric catalogs were used to study the
infrared properties of the stars. The LMC spectral type catalog
contains 1750 massive stars. A subset of 1268 of these are included in
the photometric catalog, for which uniform photometry from $0.3-24$
$\mu$m in the $UBVIJHK_{s}$+IRAC+MIPS24 bands is presented in Paper
I. The SMC spectral type catalog contains 5324 massive stars; 3654 of
these are included in the photometric catalog, for which uniform
photometry from $0.3-24$ $\mu$m is presented in Paper II. All catalogs
are available electronically.

\firstsection
\section{Active OB stars}

\subsection {O/Oe and early-B/Be stars}

In Figure 1, we plot $J_{IRSF}$ vs.  $J_{IRSF}-[3.6]$,
$J_{IRSF}-[5.8]$ and $J_{IRSF}-[8.0]$ colors for the 1967 early-B
stars from our SMC catalog, respectively, denoting their luminosity
classes, binarity and emission line classification properties by
different symbols. We compare the observed colors with colors of
plane-parallel non-LTE TLUSTY stellar atmosphere models \cite[(Lanz \&
Hubeny 2003, 2007)]{Lanz03, Lanz07} of appropriate metallicity and
effective temperatures. For reference, reddening vectors and TLUSTY
models reddened by $E(B-V)=0.2$ mag are also shown.  We clearly detect
infrared excesses from free-free emission despite not having
dereddened the stars, as in the LMC. At longer wavelengths, the excess
is larger because the flux due to free--free emission for optically
thin winds remains essentially constant with wavelength. Fewer stars
are detected at longer wavelengths because of the decreasing
sensitivity of {\it Spitzer} and the overall decline of their SEDs. We
find that the majority of early-B supergiants in the SMC exhibit lower
infrared excesses, when compared to their counterparts in the LMC, due
to their lower mass-loss rates, although certain exceptions exist and
deserve further study.

The CMDs allow us to study the frequency of Oe and Be stars, given the
low foreground and internal reddening for the SMC. Our SMC catalog
contains 4 Oe stars among 208 O stars, of which one is bluer than the
rest. There are 16 additional stars with $J_{IRSF}-[3.6]>0.5$~mag and
$J_{IRSF}<15$~mag (including all luminosity classes), whose spectra
appear normal (although the H$\alpha$ spectral region in most cases
was not observed). We refer to these as ``photometric Oe'' stars and
attribute their infrared excesses to free-free emission from a
short-lived, possibly recurrent circumstellar region, whose H$\alpha$
emission line was not detected during the spectroscopic observations
either because the gas had dispersed or because the region was
optically thick to H$\alpha$ radiation or the observation spectral
range just did not extend to H$\alpha$. Given the expectation of lower
$\dot{M}$ at SMC metallicity, we argue that such a region is more
likely to be a transient disk rather than a wind. Assuming these are
all Oe stars, we find a $10\pm2\%$ fraction of Oe stars among the O
stars in the SMC. The error in the fraction is dominated by small
number statistics. In contrast, there are 4 Oe and 14 ``photometric
Oe'' stars (with $J_{IRSF}-[3.6]>0.5$~mag and $J_{IRSF}<14.5$~mag) out
of 354 O stars in the LMC (despite the higher $\dot{M}$ at LMC
metallicity), which yields a $5\pm1\%$ fraction of Oe stars among O
stars in the LMC.

Turning to the early-B stars, the most striking feature in Figure~1 is
a distinct sequence displaced by $\sim0.8$~mag to the red. A large
fraction of the stars falling on this redder sequence have Be star
classifications, although not all Be stars reside there. Given that
the circumstellar gas disks responsible for the emission in Be stars
are known to completely vanish and reappear between spectra taken even
1 year apart (see review by \cite[Porter \& Rivinius 2003,]{Porter03a}
and references therein), the double sequence reported here provides
further evidence for the transient nature of the Be phenomenon. A
bimodal distribution at the $L-$band was previously suggested by the
study of \cite[Dougherty et al. (1994)]{Dougherty94}, which included a
sample of 144 Galactic Be stars. Our larger Be sample, which is
essentially unaffected by reddening, and the inclusion of all early-B
stars, clearly confirms the bimodal distribution. It is due to the
much larger number of Be stars classified in the SMC, in comparison to
the LMC, as well as the higher fraction of Be stars among early-B
stars in the SMC, which is $19\pm1\%$ vs.\ $4\pm1\%$ in the LMC when
considering only the spectroscopically confirmed Be stars
(cf. $\sim17\%$ for $<10$~Myr B0--5 stars; \cite[Wisniewski et
al. 2006)]{Wisniewski06}. Excluding the targeted sample of
\cite[Martayan et al. (2007a, 2007b)]{Martayan07a,Martayan07b} does
not significantly bias the statistics, since the fraction only
decreases to $15\pm1\%$. We caution that incompleteness in our
catalogs could also affect the determined fractions, if our sample
turns out not to be representative of the whole population of OB
stars.

\begin{figure}[b]
\begin{center}
 \includegraphics[width=3.6in, angle=270]{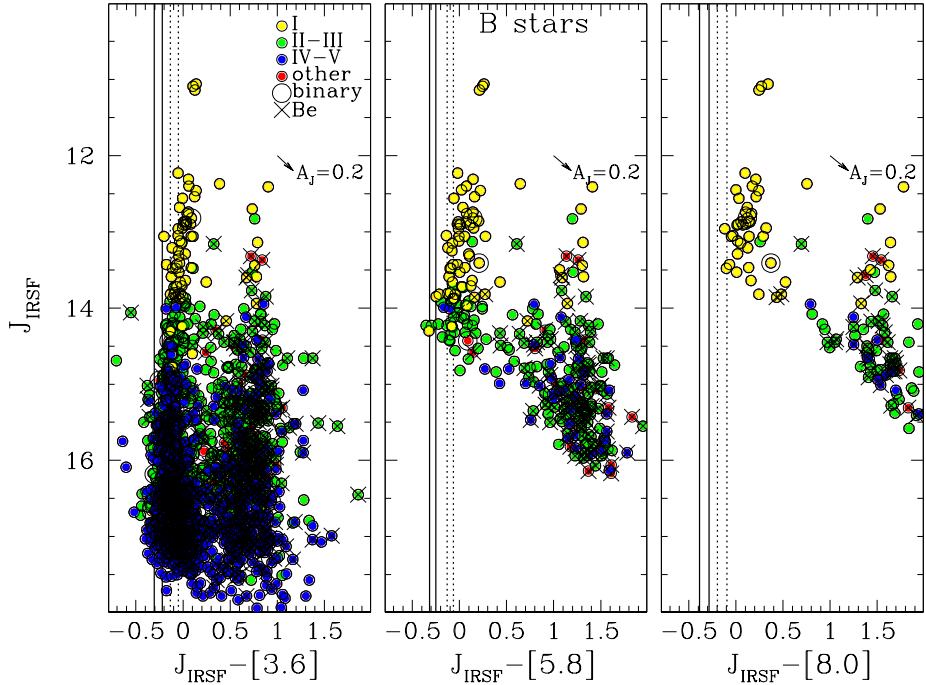} 
 \caption{Infrared excesses ($J_{IRSF}$ vs. $J_{IRSF}-[3.6]$,
   $J_{IRSF}-[5.8]$ and $J_{IRSF}-[8.0]$) for 1967 early-B stars in
   the SMC. Supergiants are shown in yellow, giants in green,
   main-sequence stars in blue, stars with uncertain classifications
   (``other'') in red, binaries with a large circle and Oe stars with
   an $\times$. The solid lines correspond to 30kK and 50kK TLUSTY
   models with $\log g = 4.0$. A reddening vector for $E(B-V)=0.2$ mag
   is shown, as well as reddened TLUSTY models by this same amount
   (dotted lines). The more luminous stars exhibit larger infrared
   excesses, which increase with $\lambda$.}
   \label{fig1}
\end{center}
\end{figure}

We proceed to define ``photometric Be'' stars as early-B type stars
with an intrinsic color $J_{IRSF}-[3.6]>0.5$~mag, given that a
circumstellar disk or envelope is required to explain such large
excesses. Including these ``photometric Be'' stars and using the same
color and magnitude cuts as for the ``photometric Oe'' stars above,
yields fractions of Be stars among early-B stars of $27\pm2\%$ for the
SMC and $16\pm2\%$ for the LMC (cf. 32\% from young SMC clusters;
\cite[Wisniewski et al. 2006)]{Wisniewski06}. We compare our results
with the fractions determined by \cite[Maeder et al. (1999)]{Maeder99}
from young clusters, i.e.\ 39\% for the SMC and 23\% for the LMC,
finding ours to be lower, although the sample selections were very
different.

These preliminary statistics (available for the first time for Oe
stars) indicate that both Oe and Be stars are twice as common in the
SMC than in the LMC. We emphasize the importance of including the
``photometric Be'' stars, which significantly increase the frequencies
of Oe/O and Be/early-B stars determined and are crucial when comparing
such stars in different galaxies. This novel method of confirming Oe
and Be star candidates from their infrared colors or a combination of
their optical and infrared colors, as recently suggested by \cite[Ita
et al. (2010)]{Ita10a} is complementary to the detailed spectroscopic
analyses by e.g.\ \cite[Negueruela et al. (2004)]{Negueruela04} on
individual Oe stars to understand their nature, although it is limited
to galaxies with low internal reddening.  We finally note that the
spectral types of Oe stars in the SMC (O7.5Ve, O7Ve, O4-7Ve and
O9-B0III-Ve) and the LMC (O9Ve (Fe II), O7:Ve, O8-9IIIne, O3e) are
earlier than those of known Galactic Oe stars, which are all found in
the O9-B0 range \cite[(Negueruela et al. 2004)]{Negueruela04}.

Finally, we note that the brightest Be stars in the SMC
($J_{IRSF}\sim13.2$~mag) are brighter than the brightest Be stars in the
LMC ($J_{IRSF}\sim13.4$~mag), i.e.\ there is a 0.7 mag difference in
absolute magnitude, given the 0.5 mag difference in the distance
moduli.

\subsubsection{Supergiant B[e] stars}

In the SMC photometric catalog, we have detected 7 luminous sources
with colors typical of sgB[e] stars (see Paper I for an introduction),
i.e.\ $M_{3.6}<-8$, $[3.6]-[4.5]>0.7$, $J-[3.6]>2$~mag. Five of these
are previously known sgB[e] stars (with R50; B2-3[e] being the
brightest in all IRAC and MIPS bands), while R4 (AzV\,16) is
classified as an LBV with a sgB[e] spectral type. In addition to
these, we find that 2dFS1804 (AFA3kF0/B[e]) has a very similar SED
(and therefore infrared colors) to the known sgB[e] 2dFS2837
(AFA5kF0/B[e]).  \cite[Evans et al. (2004)]{Evans04} also remarked on
the similarity of their spectra. We therefore confirm the supergiant
nature of 2dFS1804. The similarity of the SEDs of these sgB[e] stars,
despite the various optical spectral classifications, implies that all
are the same class of object. The cooler, composite spectral types
indicate a lower mass and perhaps a transitional stage to or from the
sgB[e] phenomenon. The only difference we find between the sgB[e]
stars in the SMC vs. the LMC is that on average they are $\sim$1-2 mag
fainter (in absolute terms).

\subsubsection{Luminous Blue Variables}

All 3 known LBVs in the SMC: R4 (AzV\,16, B0[e]LBV), R40 (AzV\,415,
A2Ia: LBV) and HD\,5980 (WN6h;LBV binary), were detected at infrared
wavelengths. R4 is the more reddened LBV, whereas the colors of
HD\,5980 (a well known eccentric eclipsing binary, see
e.g. \cite[Foellmi et al. 2008)]{Foellmi08} are similar to those of
the LBVs in the LMC. We find their SEDs to differ, given their very
different spectral types. Moreover, we find evidence for variability,
which can be confirmed from existing light curves in the All Sky
Automated Survey (ASAS) \cite[(Pojmanski 2002)]{Pojmanski02}, as
pointed out by \cite[Szczygiel et al. (2010)]{Szczygiel10}, who
studied the variability of the massive stars presented in Paper~I in
the LMC. The various SED shapes and spectral types observed depend on
the time since the last outburst event and the amount of dust formed.

\begin{discussion}

\discuss{Miroshnichenko}{What kind of positions do you have in your
  catalog of OB stars in the LMC (Spitzer, 2MASS, optical)?}

\discuss{Bonanos}{We have used the best coordinates available,
  e.g. from Brian Skiff's updated lists for the Sanduleak catalog,
  which are generally accurate to $<$1''.}

\discuss{Wisniewski}{How do you exclude or differentiate Herbig Be
  stars from classical Be stars in your data? Herbigs' transitional
  disks can show similar optical spectroscopic features: IR colors
  (especially given the dust content of the SMC/LMC) and candidate
  Herbigs have already been identified in the SMC/LMC, see e.g. Lamers
  et al. 1999; de Wit et al. 2002, 2003, 2005; Bjorkman et al. 2005.}

\discuss{Bonanos}{We have not differentiated between them, as our
  sample was selected from the literature by mainly targeting OB stars
in clusters. None of the stars in our catalog have HBe
classifications, however some could be HBe stars.} 

\end{discussion}

\end{document}